\documentclass[twocolumn,10pt,aps,prd,floatfix]{revtex4}
\usepackage{amsmath}
\usepackage{amssymb}
\usepackage{amsfonts}
\usepackage{graphicx}
\usepackage{array}
\usepackage{booktabs}
\usepackage{psfrag}

\begin{document}
\title{Inflationary Spectra from Lorentz Violating
Dissipative Models}
\author{Julian Adamek}
\email{jadamek@physik.uni-wuerzburg.de}
\author{David Campo}
\email{dcampo@astro.uni-wuerzburg.de}
\author{Jens C. Niemeyer}
\affiliation{Institut f\"ur Theoretische Physik und Astrophysik, 
         \\Universit\"at W\"urzburg,\\ 97074 W\"urzburg, Germany}
\author{Renaud Parentani}
\affiliation{Laboratoire de Physique Th\'eorique,\\ CNRS UMR 8627,
         \\Universit\'e Paris-Sud 11,\\ 91405 Orsay Cedex, France}

\begin{abstract}
The sensitivity of inflationary spectra to initial
conditions is addressed in the context of a phenomenological model
that breaks Lorentz invariance by
dissipative effects above some threshold energy $\Lambda$.
These effects are obtained dynamically by coupling the 
fluctuation modes to extra degrees of freedom which are
unobservable below $\Lambda$.
Because of the strong dissipative effects in the early propagation,
only the state of the extra degrees of freedom is relevant for 
the power spectrum. If this state is the ground state, 
and if $\Lambda$ is much larger than the Hubble scale $H$, the standard
spectrum is recovered. Using analytical and numerical methods, we
calculate the modifications for a large class of dissipative models.
For all of these, we show that the leading modification  (in an expansion in $H/\Lambda$)
is linear in the decay rate evaluated at horizon exit, and that
high frequency superimposed oscillations are not generated.
The modification is negative when the decay rate decreases
slower than the cube of $H$, which means that
there is a loss of power on the largest scales.

\end{abstract}
\maketitle

\def\pvec{{\mathbf{p}}}
\def\pphys{{P}}
\def\etaI{{\eta_p^\star}}
\def\dd{{\mathrm{d}}}
\def\GretPsi{{G^\Psi_k}}
\def\GretPhi{{G^\Phi_p}}
\def\GretPhiprime{{G^\Phi_{p'}}}

\section{Introduction}

Today's picture of cosmological evolution assumes that all 
large scale structures we observe today developed from primordial
fluctuations on top of a homogeneous and isotropic state of the early
universe. These primordial fluctuations naturally arise in the context
of inflation. As a consequence of the accelerated expansion, short
wavelength vacuum fluctuations are amplified as they exit the horizon scale.  
After the end of inflation, these fluctuations re-enter
the horizon and eventually undergo gravitational collapse.
Depending on the total number of $e$-folds of inflation, the
structures we observe today may originate from fluctuations
with extremely
small initial wavelengths as defined in the
homogeneous frame. In fact, 
unless we fine-tune the number of $e$-folds, the relevant scales 
were all well beyond the Planck scale at the onset of 
inflation~\cite{Jacobson1999}. 

Inflation, therefore, effectively acts as a spacetime
microscope, offering the tempting opportunity to probe very high
energies by looking for signatures in the primordial perturbation
spectrum~\cite{MartinBrandenberger00,JN00}. 
Let us denote the scale at which new physics becomes 
important by $\Lambda$. Depending on the nature of the dominant new
physics (which may or may not be of gravitational origin), $\Lambda$
might be the Planck scale, the string scale, or below. Demanding that
the theory yields the usual results in the infrared generically gives rise
to a suppression of the corrections by some power of $H_p/\Lambda$,
where $H_p$ is the Hubble scale at the time when the mode $p$ under
consideration left the horizon. There are thus two ways of detecting
the new physics in the perturbation spectrum. First, if the
correction contains a sharply defined phase factor which is a function
of $H_p/\Lambda$, an oscillatory
feature extending over a wide range of the power spectrum may be
produced~\cite{Danielsson,Eastheretal,NPC}. Second, even if
oscillations are absent, there is 
still a possibility that the largest scales we observe today correspond to
sufficiently large $H_p/\Lambda$ that the new physics lead to a distinctive
suppression or enhancement of the low-$l$ power spectrum at a detectable
level. While we do not find effects of the first kind in our model, we
conjecture that it generically predicts a large scale modification of
the power spectrum. 

In the absence of clear predictions from a fundamental theory, several
types of phenomenological approaches have been proposed.
If Lorentz invariance is retained at all energies, see e.g.~\cite{Kaloper}, 
there is no possibility to introduce deviations
from the standard relativistic propagation
and thus very little hope to get any signatures.
It is therefore interesting to consider the breaking of Lorentz invariance
in the ultraviolet (UV) sector, and in fact this was done in
essentially all approaches. 

In the simplest of these models, scalar or tensor perturbation modes
are prescribed to be created in their adiabatic vacuum 
at some fixed initial time, where the initial conditions are specified
either on  
a spacelike surface common to all modes~\cite{tcutoff}, 
or when the physical momentum $P = p/a$ of each mode with comoving
momentum $p$ satisfies $P = \Lambda$~\cite{Danielsson,Eastheretal,NPC}. 
These models predict a spectrum with superimposed oscillations
whose amplitude is a power of $H_p/\Lambda$ which depends 
on the degree of non-adiabaticity of the initial state.
The validity of this conclusion 
was questioned in~\cite{CampoNiemeyerParentani2007} on the basis that
the modulation  of the corrections artificially follows 
from the sharp, and thus non-adiabatic, character of 
imposing the initial state at a given instant. 

In another approach~\cite{MartinBrandenberger00,JN00}, 
Lorentz invariance is broken by introducing
deviations from the relativistic dispersion relation above
a UV scale $\Lambda$:
\begin{equation} \label{disp}
 \frac{\Omega^2}{P^2} =
 1 \pm \left( \frac{P}{\Lambda}
 \right)^n + \mathcal{O}\left(  \left(\frac{P}{\Lambda}\right)^{n+1}  \right) \, ,
\end{equation}
where $\Omega$
and $P$ are the proper frequency and the proper momentum 
as measured in the preferred frame which is assumed to
coincide with the cosmological frame~\footnote{\label{f1}
The description of dispersion or 
dissipation in empty spacetime requires to introduce a 
preferred frame. In condensed matter systems or in a heat bath, 
this frame is introduced by the medium itself.
In covariant settings, the preferred frame is 
defined by a vector field~\cite{Jacobson_OntheOriginof}. 
If this field is described dynamically, 
it aligns itself with the cosmological frame exponentially fast
in inflationary backgrounds~\cite{KannoSoda}.}.
 It has been understood that the 
standard predictions are robust~\cite{NiemeyerParentani2001},
i.e. the modifications of the 
spectra scale as a power of $H_p/\Lambda$,
provided the initial state is the asymptotic vacuum and 
the modes evolve adiabatically. 
Under these conditions, dispersive models generically predict 
no superimposed oscillations~\cite{Macher:2008yq}.

The alternative possibility that Lorentz invariance is broken by
dissipative effects has received much less attention so far.
To be realized while maintaining unitarity, 
one must introduce additional, unobservable degrees of freedom,
hereafter called $\Psi$, which couple to the 
observable field $\phi$ in the UV sector~\cite{Parentani:2007uq}.
In this paper we aim to compute the modifications of the spectrum
induced by such dissipative effects.

In order to obtain a local equation for the effective propagation
of $\phi$ after tracing out $\Psi$, a simple
class of models for the propagation
of $\Psi$ and its interaction with $\phi$ is analyzed in detail.
This class is characterized by
the ``decay rate'' $\Gamma$ of the $\phi$ modes
in the preferred frame, which 
-- in analogy to eq.~(\ref{disp}) --
we parameterize by
\begin{equation} \label{dissipparam}
  \frac{\Gamma}{P} = 
  \left(\frac{P}{\Lambda}\right)^n +
  \mathcal{O}\left(\left(\frac{P}{\Lambda}\right)^{n+1}\right)\, .
\end{equation}
If $H/\Lambda \ll 1$, 
we argue in Sec.~\ref{general} that 
the power spectrum of a wide range of dissipative
models can be effectively described by a simplified model
characterized by such a decay rate.

The paper is organized as follows. The model is presented in 
Sec.~\ref{sec:model}.
After introducing the settings in Sec.~\ref{sec:settings},
we derive the  effective equation of motion of the 
$\phi$ mode in Sec.~\ref{sec:effeom}. In Sec.~\ref{sec:2ptPhi}, the power spectrum is presented
in terms of a double integral of a noise kernel governed by $\Psi$ and the 
retarded Green function of $\phi$. 
From an approximate expression of the Green function (Sec.~\ref{sec:approxG_ret}),
we derive analytic expressions for the power spectrum 
in Sec.~\ref{sec:proppower} and~\ref{sec:highT} at
zero and high temperatures, respectively.
The numerical scheme and the results are presented in Sec.~\ref{sec:num}.

\section{Model and power spectra}
\label{sec:model}

\subsection{The model}
\label{sec:settings}

In slow-roll inflation, the background is a flat Friedmann universe
with the usual Friedmann-Lema\^itre-Robertson-Walker metric  
\begin{equation} \label{metric}
  ds^2 = - dt^2 + a^2\left(t\right) \dd \mathbf{x}^2
  = a^2\left( \eta \right) \left(  -d\eta^2 +  \dd \mathbf{x}^2 \right)\ ,
\end{equation}
and the variation of $H = \partial_t \ln a$
is governed by the slow-roll parameter $\varepsilon = - \partial_t H/H^2 \ll 1$.
Since the background is homogeneous, the fields decompose into Fourier
modes labeled by the  comoving wave vector $\pvec$.
Moreover, when the cosmological and preferred frames coincide 
(see~\cite{KannoSoda}),
the action splits into disconnected sectors, 
$S = \int \dd^3 p \, S(\pvec)$.

We consider a scalar field $\phi$. The action for the rescaled mode
$\Phi_\pvec = a \phi_\pvec$ is 
\begin{equation} \label{S_Phi}
  S_\Phi\left(\pvec\right) =
  \int \dd \eta~\Phi_\pvec^\dagger \left[-\partial_\eta^2 - 
   \omega_p^2 \right] \Phi_\pvec~,
\end{equation}
where the conformal frequency is given by
\begin{equation}
  \omega_p^2\left(\eta\right) = p^2 - \frac{\partial_\eta^2 f}{ f}\, .
\end{equation}
For the scalar field $\phi$, one has $f = a$. The same is true 
for tensor modes, whereas for density perturbation modes 
$f =\sqrt{\varepsilon} a$. 
Given the simplicity of these substitutions, we
limit our discussion to the scalar field in this work.

The power spectrum of $\phi$ is given 
by the Fourier transform of the two-point correlation function
at equal times
\begin{eqnarray} \label{powerspectrum}
   \mathcal{P}_p\left(\eta\right)  &\equiv& 
   \int\!\!\frac{\dd^3\mathbf{x}}{(2 \pi)^3} \,  e^{i \pvec \mathbf{x}} \, 
   \bigl\langle \phi 
   \left(\eta, \mathbf{x}\right) \phi
   \left(\eta, \mathbf{0}\right) \bigr\rangle\,\nonumber\\
   &=&\frac{1}{2 a^2(\eta)} \int\!\dd^3\pvec' \, 
   \bigl\langle \bigl\lbrace\Phi_\pvec 
   \left(\eta\right), \Phi_{\pvec'}
   \left(\eta\right)\bigr\rbrace \bigr\rangle\, ,
\end{eqnarray}
where $\langle\,\cdot\,\rangle$ and $\lbrace\, \cdot\,,\,\cdot\,\rbrace$ denote
the quantum expectation value and the anticommutator, respectively.

In the Bunch-Davies vacuum 
(the adiabatic vacuum for $P= p/a \to \infty$),
the power spectrum is simply 
\begin{equation} \label{BDpower}
   \mathcal{P}_p(\eta) = \frac{1}{a^2(\eta)}
   \left| \Phi_\pvec^{\rm in}(\eta)\right|^2  
  \, .
\end{equation}
Here, $\Phi_\pvec^{\rm in}$ denotes
the unit Wronskian positive frequency solution of
$\left[\partial_\eta^2 + \omega_p^2  \right] \Phi_\pvec =0$.
Evaluated at late time (${p}/{a} \ll H_p$),
one obtains the standard expression
\begin{equation} \label{standardpower}
   \mathcal{P}_p^0 = \frac{H_p^2}{2p^3}\, ,
\end{equation}
where $H_p$ is the value of $H$ when the $\pvec$-mode
exits the Hubble scale. 

The goal of this paper is to compute (numerically where necessary) the
modifications of this power spectrum 
due to interactions with some additional field $\Psi$ 
inducing dissipative effects as parametrized in eq.~(\ref{dissipparam}).
To this end, we use the model introduced in~\cite{Parentani:2007uq}
whose essential feature is that $S_\Psi + S_{\mathrm{int}}$,
the action of $\Psi$ plus that
governing the $\Phi$-$\Psi$ interactions,
breaks Lorentz invariance in the UV sector. As a result, 
similarly to the dispersive models of eq.~(\ref{disp}),
the propagation of $\Phi$ remains unaffected in the low-energy sector,
whereas it is no longer Lorentz invariant in the UV, even in the
vacuum.

There is of course a lot of freedom to choose $S_\Psi$ and 
$S_{\mathrm{int}}$. But as explained in the Introduction, our aim 
 is to obtain simple equations for the effective 
propagation of $\Phi$ after having traced over $\Psi$.
From this point of view, $\Psi$ is introduced only to give rise to
dissipative effects while preserving unitarity.
Since Gaussian models are the simplest and yet do the job,
we work with quadratic actions.
For further discussion concerning the general character of these actions,
see~\cite{Parentani:2007uq} and Sec.~\ref{general} below.

Because of Gaussianity, the action still splits 
into disconnected sectors:  
\begin{equation}
\label{sum3}
  S = 
  \frac{1}{2} \int \dd^3 \pvec~\left(S_\Phi\left(\pvec\right) +
    S_\Psi\left(\pvec\right) +  
  S_{\mathrm{int}}\left(\pvec\right)\right)~.
\end{equation}
The action $S_\Phi$ is given in eq.~(\ref{S_Phi}),
and we use
\begin{eqnarray} \label{S_Psi}
  S_\Psi\left(\pvec\right) &=& 
  \int \dd t \int \dd k~\Psi_{\pvec, k}^\dagger 
  \left[-\partial_t^2 - (\pi \Lambda k)^2\right] \Psi_{\pvec, k}\,, \quad
\\ 
 \label{S_int}
  S_{\mathrm{int}}\left(\pvec\right) &=& 
  \int \dd \eta~g_p\left(\eta\right) 
  \int \dd k \left(\Phi_\pvec \partial_\eta \Psi_{\pvec, k}^\dagger 
  + {\rm h.c.}\right)\, .
\end{eqnarray}
In the action~(\ref{S_Psi}), the proper frequency of the $\Psi_k$ 
is dimensionalized by $\Lambda$ which is the only constant (proper) scale. 
These frequencies are chosen to remain constant as the universe expands
because this guarantees that the $\Psi_k$  are not excited by the
cosmological expansion. Moreover, the $\Psi_k$ carry no spatial
momentum. They are thus at rest with respect to the cosmological
frame.

In eq.~(\ref{S_int}), the coupling is bilinear, so that the model is 
indeed Gaussian. Hence, we can integrate out the $\Psi_k$. Moreover, the index
$k$ is chosen to be continuous so that the Poincar\'{e} recurrence time of
the system is infinite in order to effectively obtain dissipation
\footnote{In this paper, we refrain from giving any specific
interpretation to the $\Psi_k$ fields 
since our aim is to determine the consequences of dissipation
as opposed to justify dissipation from first principles.
Nevertheless, if one wishes, the parameter $k$ may be 
viewed as a momentum in a flat fourth spacelike direction. 
One can then draw a connection to the five-dimensional 
models described in \cite{LibanovRubakov2005}, 
where the equivalent of the $\Psi_k$ propagate in the bulk.}.
The time dependent coupling $g_p$ 
will be chosen so as to produce the desired dependence on $P/\Lambda$
in the decay rate $\Gamma$, see Sec.~\ref{sec:effeom}.
In order to obtain an effective equation of motion
of $\Phi$ which is local in time, see eq.~(\ref{phideq}) below,
we have used a derivative coupling.

\subsection{The effective equation of motion of $\Phi$}
\label{sec:effeom}

Since our model is Gaussian, all equations of motion are linear 
in the field amplitude. Hence, they can be treated as equations
for the field operators in the Heisenberg picture.

The equation for $\Psi_k$ is
\begin{equation}
\label{eomPsi}
   \left[\partial_t^2 + \Omega_k^2\right] \Psi_{\pvec, k}
   ~=~-\partial_t \left(g_p \Phi_\pvec\right)~,
\end{equation}
where $\Omega_k~\equiv~\pi \Lambda \left|k\right|$.
Its general solution is 
\begin{equation} \label{psisolution}
   \Psi_{\pvec, k}(t)~=~ 
   \Psi_{\mathbf{p}, k}^0(t) - \int \dd t'~\GretPsi(t, t')~ 
   \partial_{t'}\left(g_p \Phi_\pvec\right)~.
\end{equation}
Here, $\Psi_{\mathbf{p}, k}^0$ obeys the homogeneous 
equation, and the retarded Green function is given as
\begin{equation}
   \GretPsi(t, t')~=~  
   \int\frac{\dd \Omega}{2\pi}~ \frac{e^{-i \Omega \left(t - t'\right)}}{
    \Omega_k^2 - \Omega^2 - i \varepsilon \Omega} ~.
\end{equation}

Similarly, the equation for $\Phi$ reads
\begin{equation}
   \left[\partial_\eta^2 + \omega_p^2\left(\eta\right)\right] \Phi_\pvec~=~
   \int \dd k~g_p \partial_\eta \Psi_{\pvec, k}~.
\end{equation}
Inserting eq.~(\ref{psisolution}) into the r.h.s.~yields
\begin{multline}
   \left[\partial_\eta^2 + \omega_p^2\left(\eta\right)\right] \Phi_\pvec
   =\int \dd k~g_p \partial_\eta \Psi_{\mathbf{p}, k}^{0}~- \frac{g_p}{\Lambda} 
   \partial_\eta \left(g_p \Phi_\pvec\right)~,
\label{eom1}
\end{multline}
where we have used 
\begin{equation}
\int \dd k~\partial_{t'} \GretPsi(t, t') = 
- \frac{\delta\left(t - t'\right)}{\Lambda}\, .
\end{equation}
It is this equation which has motivated our choice of 
the action~(\ref{S_Psi}) and~(\ref{S_int}).
Indeed, in general one would have obtained a non-local equation, 
whereas here, the effective equation of motion 
of $\Phi_{\bf p}$ is simply
\begin{equation} \label{phideq}
    \left[\partial_\eta^2 + 2 \gamma_p \partial_\eta + 
    \omega_p^2 + \partial_\eta \gamma_p\right] \Phi_\pvec ~=~ 
     g_p \int \dd k~\partial_\eta \Psi_{\mathbf{p}, k}^{0}\, .
\end{equation}

The term $2 \gamma_p  \partial_\eta $ gives rise to  
dissipative effects. They are governed by the 
decay rate (in conformal time) 
\begin{equation} \label{gamma}
   \gamma_p \left(\eta\right)  = 
   \frac{\left(g_p(\eta)\right)^2}{2 \Lambda}\, .
\end{equation} 
The time dependence of the coupling $g_p$
is fixed by the following conditions. 
We first demand that the scale $\Lambda$ be the 
{\it proper} energy at which interactions
between $\Phi$ and $\Psi_k$ 
appear, irrespective of the comoving momentum label $p$. We also 
impose $g_p \rightarrow 0$ for low momenta $\pphys \ll \Lambda$, 
so that the  $\Phi_{\bf p}$ decouple from $\Psi$ 
and propagate freely. This is required by particle physics 
observations which put severe constraints on possible violations
of Lorentz invariance, see e.g.~\cite{Jacobson:2005bg}.  

In analogy to eq.~(\ref{disp}), in order to cover the general case 
we classify dissipative effects according to the lowest order of
$\pphys / \Lambda$: 
\begin{equation} \label{gammaseries}
   \frac{\gamma_p}{p}= \frac{\Gamma}{P}=
   \left(\frac{\pphys}{\Lambda}\right)^n
\left( 1+ \mathcal{O}\left(\frac{\pphys}{\Lambda}\right) 
\right)\, .
\end{equation}
The first series coefficient can always be set to unity by a 
redefinition of $\Lambda$. From this equation one can already conclude 
that in cosmology, as the proper momentum 
$P = p/a$ redshifts, the modes go from a
strongly dissipative regime
$\Gamma/P =\mathcal{O}(1)$
for $P \gtrsim \Lambda$,
to an underdamped regime
where  $\Gamma/P \ll 1$. Using eq.~(\ref{gamma}), we see that 
the $n$-th coupling function should be taken as
\begin{equation}
  g_p= \sqrt{2 p \Lambda}\left(\frac{P}{\Lambda}  \right)^{n/2} 
     = \sqrt{2 p \Lambda} \left(\frac{p}{a\Lambda} \right)^{n/2} \, .
\end{equation}

In this we follow the same approach as previously employed
in dispersive models. First, we replace the relativistic relation by an 
effective equation of motion, eq.~(\ref{phideq}), where the dissipative effects 
are chosen from the outset, and second, we
determine the modifications of the power spectrum 
induced by this replacement. In this paper we have no ambition
to put forward ``privileged'' (or ``inspired'') dispersive/dissipative models
that could be derived from first principles.
Yet another way to position this approach is to state that we follow a
bottom-up rather than a top-down route to new physics.

\subsection{The power spectrum in dissipative settings}
\label{sec:2ptPhi}

The general solution of eq.~(\ref{phideq}) is 
\begin{equation} \label{soleffEOM}
   \Phi_{\pvec}(\eta) = \Phi_{\pvec}^d(\eta) 
   + \int\!\!d\eta' \, \GretPhi(\eta, \eta') 
 \int\!\! \dd k\,  g_p \, \partial_{\eta'} \Psi_{\mathbf{p}, k}^{0}
 \, ,
\end{equation}
where $\Phi_{\pvec}^d(\eta)$ and $\GretPhi$ are
the homogeneous solution and the retarded Green function, respectively . 

The homogeneous solution $\Phi_{\pvec}^d$ decays as
\begin{equation} \label{hom}
  \Phi_{\pvec}^d(\eta) \propto 
   \exp\left( -\int^{\eta}_{\eta_{\rm in}}\!\!\dd\eta' \gamma_p\left(\eta'\right) 
  \right) \, ,
\end{equation}
where the initial time $\eta_{\rm in}$ fixes the moment when
$\Phi$ and $\Psi$ start to interact. Since we do not want to fine-tune
the number of  $e$-folds, we assume that $\eta_{\rm in}$ 
is located deep in the ultraviolet regime
\begin{equation} \label{eta_0}
  \frac{P_{\rm in}}{\Lambda}= \frac{p}{a(\eta_{\rm in})\Lambda} \gg 1 \,.
\end{equation}
In this case, the $\Phi_{\pvec}^d$
does not contribute to any observable at late time, 
implying that  only the state of the $\Psi_k$ is relevant \footnote{The situation
is a bit more subtle when there is an overdamped regime $(\gamma_p^2 > \omega_p^2)$
in the UV, since the decay of the homogeneous solution may slow down considerably.
This issue is addressed in detail in Sec.~\ref{sec:coupling}.}. 
Therefore, the power spectrum of super-horizon modes is insensitive to
the initial state of $\Phi_\pvec$.

Let us establish this important property in more detail. 
If the initial state factorizes, 
i.e.~$\Psi_{\mathbf{p}, k}^{0}$ and $\Phi_{\pvec}^d$ are not initially correlated,
the anti-commutator of $\Phi$ reads 
\begin{multline} 
\label{fullG_a}
   \bigl\langle 
  \bigl\{ \Phi_{\pvec}(\eta),\, \Phi_{\pvec'}(\eta') \bigr\} \bigr\rangle 
  = \bigl\langle 
  \bigl\{ \Phi_{\pvec}^d(\eta),\, \Phi_{\pvec'}^d(\eta') \bigr\} \bigr\rangle
\\
   + \int_{\eta_{\rm in}}^{\eta}\!\!\dd\eta_1 \int_{\eta_{\rm in}}^{\eta'}\!\!\dd\eta_2 \, 
   \GretPhi(\eta, \eta_1)
   \GretPhiprime(\eta', \eta_2) \mathcal{N}_{\pvec, \pvec'}(\eta_1,\eta_2)\, ,
\end{multline}
where we introduced the so-called noise kernel
\begin{multline} \label{Nkernel}
  \mathcal{N}_{\pvec, \pvec'}(\eta_1,\eta_2) ~\equiv~ \delta^3\left(\pvec - \pvec'\right) N_p\left(\eta_1, \eta_2\right) ~=~ 
   \\g_p\left(\eta_1\right) g_p\left(\eta_2\right)
   \int \dd k \dd k' \bigl\langle 
   \bigl\{ \partial_{\eta_1} \Psi_{\mathbf{p}, k}^{0}, 
  \partial_{\eta_2} \Psi_{\mathbf{p'}, k'}^{0 \dagger}   \bigr\} \bigr\rangle~,
\end{multline}
whose properties will be specified below.
We have simply assumed that the state of $\Psi_k$ is homogeneous. 
Because of the decay given in eq.~(\ref{hom}), 
we immediately conclude that the first term in eq.~(\ref{fullG_a})
will be exponentially damped. Hence, the anti-commutator, and thus
the power spectrum~(\ref{powerspectrum}), are entirely given by the term
which is driven by the noise kernel.

In other words, if inflation lasts long enough 
and if dissipation is sufficiently efficient in the UV
(this requirement will be discussed in more detail below),
the equation
\begin{multline} \label{G_a}
   \mathcal{P}_p = \lim_{\eta_{\rm fin} \rightarrow 0^-} \frac{1}{2 a^2\left(\eta_{\rm fin}\right)}
\\  
   \times \int_{\eta_{\rm in}}^{\eta_{\rm fin}}\!\!\dd\eta_1 \int_{\eta_{\rm in}}^{\eta_{\rm fin}}\!\!\dd\eta_2 \,  
   \GretPhi(\eta_{\rm fin}, \eta_1)\, 
   \GretPhi(\eta_{\rm fin}, \eta_2)\,  N_p(\eta_1,\eta_2)\, ,
\end{multline}
is exact, and we may furthermore take $\eta_{\rm in}$ to $-\infty$.
This equation replaces the usual expression of eq.~(\ref{BDpower}),
governed by the norm of the free mode $\Phi^{\rm in}_\pvec$, and valid
both for relativistic and modified dispersion relations.

In view of eq.~(\ref{G_a}), we see  
that dissipation
affects the structure of the equations much more profoundly
than dispersion does. 
We also understand
that the introduction of $\Psi$ could not  
have been avoided, since 
 $\Psi$ determines {\it both} 
the noise kernel $N_p$ (through its anti-commutator) and the decay rate $\gamma_p$
(through its retarded Green function, see eqs.~(\ref{psisolution},~\ref{eom1})).
These must be related to each other by a fluctuation-dissipation
relation, see~\cite{Parentani:2007uq} for a brief review in the present context.
This explains why, unlike dispersion, one cannot treat
dissipative effects by simply introducing an imaginary term in the dispersion relation.

The remarkable property of dissipation 
when it is introduced by coupling to 
some dynamical degrees of freedom
is that $\mathcal{P}_p$ is independent of all their properties
if $H_p \ll \Lambda$ and if they 
are in their ground state. 
Moreover, in this case, as we will 
show, the spectral power~(\ref{G_a}) 
agrees with the 
standard value given by eq.~(\ref{standardpower}).

\subsection{The noise kernel}
\label{sec:noise}

The definition of the model is complete 
once we specify the state of the $\Psi_k$.
For simplicity, we only consider thermal states. 
Recall that the proper frequencies of $\Psi_k$ are time independent
so that the proper Hamiltonian of $\Psi_k$ has stationary eigenstates.

Then, at temperature $T$, the noise kernel~(\ref{Nkernel}) is 
\begin{multline}
\label{Nexplicit}
   N_p = 
   g_p\left(\eta_1\right) g_p\left(\eta_2\right) 
   a\left(\eta_1\right) a\left(\eta_2\right)
\\   \times 
  \frac{2 T}{\Lambda} \partial_{t_1}\mathrm{coth}
   \left(\pi T \left(t_1 - t_2\right)\right)~.
\end{multline}
This directly follows from the fact that  
the (free) fields can be decomposed as
\begin{equation}
\label{modeexpansion}
   \Psi_{\pvec, k}^0(t)~=~ 
    \mathrm{\hat{a}}_{\pvec, k}\,  \psi_k(t) + 
    \mathrm{\hat{a}}_{-\pvec, -k}^\dagger\,  \psi_k^\ast(t)\, ,
\end{equation}
where $\mathrm{\hat{a}}_{\pvec, k}$ satisfy canonical commutation
relations and where the isotropic mode functions
\begin{equation} \label{minkowskimode}
   \psi_k~=~ 
   \frac{1}{\sqrt{2 \Omega_k}} e^{-i \Omega_k t} \, ,
\end{equation}
have unit Wronskian $W[\psi_k] \equiv 2 \mathrm{Im}(\psi_k \partial_t \psi_k^\ast) = 1$.
We have also used the fact that in 
the thermal states one has 
\begin{equation}
  2 \bigl\langle \mathrm{\hat{a}}_{\pvec, k}^\dagger \mathrm{\hat{a}}_{\pvec, k} 
  \bigl\rangle_T + 1~=~\mathrm{coth}\left(\frac{\Omega_k}{2 T}\right)~.
\end{equation}

At high temperature, as usual, the kernel becomes local 
\begin{equation}\label{NhighT}
   \lim_{T \rightarrow \infty}
   N_p = 4T \, \gamma_p(\eta_1) \, a(\eta_1) \,  \delta(\eta_1 - \eta_2)\, ,
\end{equation}
where we used 
$\delta(t_1 - t_2) = \delta(\eta_1 -\eta_2) / a(\eta_1)$ 
and eq.~(\ref{gamma}). 
At zero temperature (vacuum) instead, one gets
\begin{equation} \label{NzeroT}
\lim_{T \rightarrow 0}
   N_p = \frac{4}{\pi} 
   \sqrt{\gamma_p(\eta_1) \gamma_p(\eta_2)} \, 
   a(\eta_1) a(\eta_2)\, 
   \partial_{t_1} 
   \frac{{\rm PV}}{\left(t_1 - t_2\right)}\, ,
\end{equation}
where the singular behavior should be interpreted as 
the derivative of the Cauchy principal value.

\subsection{Retarded Green function}
\label{sec:G_ret}

To compute the power spectrum~(\ref{G_a}), we need the
retarded Green function of eq.~(\ref{phideq}). 
It satisfies the boundary conditions
\begin{equation} \label{boundaryGret}
  \GretPhi\left(\eta'=\eta \right) = 0\, ,\quad
  \partial_{\eta}\GretPhi \vert_{\eta'=\eta} = 1\, .
\end{equation}
Therefore, it can be written as 
\begin{equation}
\label{gretansatz}
   \GretPhi\left(\eta, \eta'\right)~=~- 
   \theta\left(\eta - \eta'\right) 
   \frac{2 \mathrm{Im}\left(
     \varphi_p\left(\eta   \right) \varphi_p^\ast\left(\eta'\right)\right)}{\left. 
   W\left[\varphi_p\right]\right|_{\eta'}}~,
\end{equation}
where the mode function 
$\varphi_p\left(\eta\right)$ may be any homogeneous solution 
of eq.~(\ref{phideq})
that has a nondegenerate Wronskian $W\left[\varphi_p\right]$. 

We introduce the function
\begin{equation} \label{visibility}
\mathcal{I}_p\left(\eta, \eta_0\right)~\equiv~
\int_{\eta_0}^\eta \gamma_p\left(\eta'\right) \dd \eta'\, ,
\end{equation}
which gives the amount of dissipation
from $\eta_0$ to $\eta$.
It will play a crucial role in what follows.
Using it, we can get rid of the 
friction term in eq.~(\ref{phideq}),
by writing 
\begin{equation} \label{Liouville}
\varphi_p\left(\eta\right)~=~e^{-\mathcal{I}(\eta, \eta_0)}\, 
\chi_p\left(\eta\right)~.
\end{equation}
Indeed, $\chi_p$ obeys
\begin{equation} \label{chieqom}
\left[\partial_\eta^2 + \omega_p^2\left(\eta\right) - \gamma_p^2\left(\eta\right)\right] \chi_p\left(\eta\right)~=~0~.
\end{equation}
Taking into account the time dependence of the Wronskian
of $\varphi_p$, 
eq.~(\ref{gretansatz}) can be rewritten as 
\begin{multline} \label{gretexplicit}
    \GretPhi(\eta, \eta')= 
    -2\theta(\eta - \eta') 
   \mathrm{Im}\left[\chi_p(\eta) \chi_p^\ast(\eta')\right]
   e^{- \mathcal{I}_p(\eta, \eta')}
   \, ,
\end{multline}
when the constant Wronskian of $\chi_p$ was chosen to unity.
It should be noticed that only the decay accumulated from 
$\eta$ to $\eta'$ appears in $ \GretPhi$.  
The fact that $\eta_0$ must drop out can be seen from eq.~(\ref{boundaryGret}).
In fact, the second equality replaces the equal time 
commutation relation in the presence of interactions,
see \cite{Parentani:2007uq} for further details.

Returning to eq.~(\ref{G_a}), the presence of the two functions 
$\mathcal{I}_p$ evaluated both until $\eta_{\rm fin} = 0$ 
limits the past history that is relevant to the
power spectrum of super-horizon modes. To characterize this
relevant domain, 
we define the time $\etaI$ by the moment where 
\begin{equation} \label{Icrossing}
  \mathcal{I}_p\left(\eta_{\rm fin} = 0, \etaI \right)~=~1
   \, .
\end{equation}
Times earlier as $\etaI$ play no significant 
role in the power spectrum.
In other words, $\mathcal{I}_p$ can be considered as an ``optical depth''.

Having established these features, we can now  
explain why, if $H \ll \Lambda$, any dissipative model 
exhibiting dissipation above $\Lambda$ behaves as if it belonged to 
the class of models we just considered.

\subsection{General properties of dissipative models}
\label{general}

The models we studied are based on several simplifying assumptions.
First, they are Gaussian;  second, the  
frequency of the $\Psi_k$ is constant; and third, 
a derivative was introduced in the action $S_{\mathrm{int}}$ in order to get a
local equation for $\Phi$. 
Nevertheless, the features we obtained are more general:
they will be found 
in all dissipative models respecting minimal assumptions that we now clarify.

Before listing these conditions,
it should be noticed that when dealing with nonlinear interactions, 
it is no longer convenient to work with the mode operator $\Phi_\pvec$
as we just did. Instead, it is appropriate to study the effective evolution
in terms of the two-point correlation functions of $\Phi_\pvec$, 
see~\cite{Parentani:2007uq}. 
In particular, it can be shown that the expectation value of the anti-commutator of $\Phi$,
the l.h.s.~of eq.~(\ref{fullG_a}),
always obeys a {\it linear} integro-differential equation 
with a source~\cite{Bergesreview}. This also applies to
non-derivative, bilinear couplings, so that
the following discussion includes both cases.

Adopting this language, we can transpose the two conditions we
used in Section \ref{sec:2ptPhi}.
First, the dissipative effects should be strong enough
so as to erase the contribution of the homogeneous solution of this 
integro-differential equation. In this we recover, in the language of 
two-point correlation functions, the neglect of the homogeneous solution
$\Phi^d_\pvec$ of eq. (\ref{phideq}).
Second, the state of the entire system must be spatially homogeneous.
When both conditions are met,
the expectation value of the anti-commutator of the Fourier mode $\Phi_\pvec$ 
is driven by a $p$-dependent (c-number) source
through the above-mentioned integro-differential equation. 
This implies that the power spectrum will be  given by eq.~(\ref{G_a})
in {\it any} (unitary) dissipative model, Gaussian or not.
In other words, the power spectrum is always governed by a
${p}$-dependent kernel $N_p$ and a retarded Green function $\GretPhi$.

Let us begin with the kernel $N_p$, which
encodes the properties of the state of the system.
In Gaussian models, it is simply given by the expectation value of the anti-commutator
of the r.h.s.~of eq.~(\ref{phideq}).
In non-Gaussian models, it must be computed order by order in a loop expansion. 
This calculation might turn out to be difficult, 
but (in renormalizable theories) $N_p$ is a well defined kernel 
which is given by the real part of the 
(renormalized and time-ordered) self-energy of $\Phi_\pvec$
(see for instance Appendix B in \cite{Campo:2008ij}).
Therefore, when $N_p$ has been computed, it will ``drive'' the power spectrum
as indicated in eq.~(\ref{G_a}).

Let us briefly discuss the modifications one encounters
when the proper frequencies $\Omega_k(t)$ of the
$\Psi_k$ depend on time. In this case, their state 
will be parametrically excited. 
However, if the variation of $\Omega_k(t)$ is slow enough, this
amplification will
be exponentially suppressed by virtue of the adiabatic theorem.
(A similar situation is expected when dealing with
non-Gaussian models.)
Then, if the $\Psi_k$
are initially in (or close to) their ground state,
$N_p$ will essentially
be the noise of the adiabatic vacuum of the $\Psi_k$.

Let us now turn to the effects of dissipation.
In a general model, one would lose 
the local character of eq.~(\ref{phideq}).
However, in all models (Gaussian or not),
the retarded Green function of $\Phi$
obeys a {\it linear} integro-differential equation of the form 
\begin{multline} \label{nonloc}
  \left[\partial_\eta^2 + \omega_p^2 \right]\GretPhi(\eta, \eta') +   
  \int^{\eta}\!\! d\eta_1 \, {\cal D}_p(\eta, \eta_1) \, \GretPhi(\eta_1, \eta')\\ = 
  \delta(\eta - \eta')\, ,
\end{multline} 
where the non-local kernel ${\cal D}_p$ 
generalizes what we had in eq.~(\ref{phideq}) in that,
when 
${\cal D}_p = \partial_{\eta_1}\delta(\eta -\eta_1) 2\gamma_p$,
one recovers the usual odd term of that equation. The kernel
${\cal D}_p$ is antisymmetric in the exchange of its arguments and 
describes dissipation. 
It is related to the imaginary part of the time-ordered
self-energy, as $N_p$ was related to the real part, and 
is therefore also well-defined and computable, at least perturbatively.
Moreover, as for $N_p$, if the state of the $\Psi_k$ evolves adiabatically, 
${\cal D}_p$ is the dissipation kernel in the adiabatic vacuum.

Concerning the power spectrum, 
we saw in Sec.~\ref{sec:G_ret} that
in expanding universes with $H / \Lambda \ll 1$,
only the evolution in the underdamped regime is relevant.
In this low-energy, weakly dissipative regime, 
the non-local equation~(\ref{nonloc}) can be approximated by a local one
(i.e. similar to eq.~(\ref{phideq})
with an effective damping rate $\gamma_{\rm eff}$)
provided 
the characteristic comoving time describing the retardation 
effects of ${\cal D}_p$ is much smaller than $\omega_p^{-1}$, or equivalently
that the corresponding cosmological time is much smaller than $H_p^{-1}$.This approximation is similar to the diffusion approximation in kinetic theory.
In this case, we can approximate
\begin{equation}\label{expD}
  \int^{\eta}\!\! d\eta_1 \, {\cal D}_p(\eta, \eta_1) \, \varphi(\eta_1) 
   \simeq 2 \gamma_{\rm eff}(\eta) \partial_\eta \varphi(\eta)
\, ,
\end{equation}
where $\gamma_{\rm eff}(\eta)$
depends in general on $\eta$
and on the state of the system. 
(In the case one would consider the same model
in Minkowski spacetime, and in its ground state, the above equation is easily obtained in the frequency representation by performing a Taylor expansion in the frequency
and truncating at first order.
In that case, $\gamma_{\rm eff}$ would be constant. 
In an expanding universe, it becomes time dependent through the scale factor $a(t)$.)

In conclusion, a general $\Phi$-$\Psi$ model 
where 
(i) $H / \Lambda \ll 1$, 
(ii) the state of $\Psi_k$ evolves adiabatically, 
(iii) the characteristic time of ${\cal D}_p$ is much smaller than $H_p^{-1}$
so that eq.~(\ref{expD}) holds, 
will give the 
same power spectrum as that of the corresponding simplified model 
governed by eqs.~(\ref{sum3}-\ref{S_int})
with the coupling $g$ matching the 
effective decay rate $\gamma_{\rm eff}$
through eq.~(\ref{gamma}).

\section{Analytical treatment}
\label{sec:analytic}

We present some analytical expressions for the power spectrum which will
facilitate the interpretation of the numerical results.
They are valid for $H \ll \Lambda$ and in the slow-roll regime.

\subsection{More properties of the retarded Green function}
\label{sec:approxG_ret}

As we are interested in the power spectrum of superhorizon modes, 
we can make a first approximation by factorizing the growing mode of 
$\chi_p(\eta)$:
\begin{equation} \label{growingmode1}
  \chi_{p} \simeq \frac{i H_p \, a}{\sqrt{2 p^3}}~.
\end{equation}
By eq.~(\ref{gammaseries}), this is a solution of eq.~(\ref{chieqom}) for 
$\left|p \eta\right| \ll 1$. 
Hence, for $\eta_{\rm fin} \to 0^-$ we have
\begin{multline} \label{growingmode2} 
  \GretPhi\left(\eta_{\rm fin} \to 0^-, \eta\right)~\simeq~\theta(-\eta) \frac{2 H_p \, a(\eta_{\rm fin})}{\sqrt{2 p^3}}\\\times
  \mathrm{Re}\left[ \chi_p(\eta)  \right] 
  ~e^{- \mathcal{I}_p(\eta_{\rm fin}, \eta)}~.
\end{multline}
$\GretPhi$ (as function of $\eta$)
oscillates with a slowly varying envelope.

We now give an analytic approximation to eq.~(\ref{growingmode2})  
valid in the slow-roll approximation and 
in the case of scale separation: 
\begin{equation} \label{approxcond}
   \varepsilon ,\,  \frac{H}{\Lambda} \ll 1\, .
\end{equation}
Let us first give an approximation for the envelope, given by
the exponential in eq.~(\ref{growingmode2}).
We take $\gamma_p$ to be of the form~(\ref{gammaseries}), 
i.e.~$\gamma_p \propto a^{-n}$ for modes below the UV scale $\Lambda$.
The term coming from the upper bound is then negligible since
$\gamma_p(\eta) \to 0$ for $\eta \to 0$.
Hence during inflation, the integral is dominated by the lower bound, 
and we may thus estimate
\begin{eqnarray} \label{approxGretdamping}
  \mathcal{I}_p(\eta_{\rm fin}, \eta)
  &=& 
  \int_{a(\eta)}^{a(\eta_{\rm fin})} \frac{\gamma_p}{H a^2} \dd a
  \nonumber \\
  &\simeq& \frac{1}{n + 1} \frac{\gamma_p(\eta)}{H(\eta) a(\eta)}~,
\end{eqnarray}
where we have used the slow-roll approximation.

Let us now consider the term 
$\mathrm{Re}\left[ \chi_p\left(\eta\right)  \right] $.
The equation of motion, eq.~(\ref{chieqom}),
has an oscillating solution inside the horizon. However, the
term $\gamma_p^2$ in the effective squared frequency, i.e.~the
frequency shift due to dissipation, introduces some non-adiabaticity
to the evolution of the mode close to the time when it leaves the
ultraviolet regime. This implies that the retarded Green function
receives non-adiabatic corrections for very early times and may even
stop oscillating in the case where there is an overdamped regime
($\gamma_p^2~>~\omega_p^2$) in the UV.
However, in the case of scale
separation this ultraviolet
behavior of the mode function occurs 
only where the envelope is exponentially small  
by a factor $e^{-\mathcal{O}\left(\Lambda / H\right)}$. 
In other words, the dispersive effects 
induced by dissipation
are damped by dissipation itself (see also
fig.~\ref{fig:nonadiabaticity-squar}). 

Finally, notice that the noise kernel~(\ref{Nexplicit}) is
proportional to $\gamma_p$ 
and therefore vanishes as $\eta_1, \eta_2 \to 0^-$. 
In conclusion, the double integral~(\ref{G_a}) 
takes its value in the vicinity of
$\eta_1, \eta_2 \simeq \etaI$ 
defined by eq.~(\ref{Icrossing}). 
If $H \ll \Lambda$, one finds that $\chi_p(\etaI)$
is well inside the horizon and at the same time sufficiently below the
UV regime, such that it is justified to use a unit Wronskian free oscillator in
place of $\chi_p$ in order to estimate the power spectrum. In other
words, as an approximation, we set 
\begin{equation}
   \mathrm{Re}\left[ \chi_p(\eta \sim \etaI)  \right]~\approx~ 
   \frac{1}{\sqrt{2 p}} \cos(p \eta)~,
\end{equation}
and thus
\begin{multline} \label{approxGret}
   \GretPhi(\eta_{\rm fin} \to 0^-, \eta \sim \etaI) ~\simeq~ 
   \theta(-\eta) \frac{H_p a(\eta_{\rm fin})}{p^2}  \\
   \times \cos(p \eta) e^{-\frac{1}{n + 1} \frac{\gamma_p(\eta)}{H(\eta) a(\eta)}}
\end{multline}
when we evaluate the double integral. It should be stressed that this
is only done to get an analytic estimate, and that no approximation is
required for the numerical treatment since we can always solve
eq.~(\ref{chieqom}) numerically.

\subsection{General properties of the power spectrum}
\label{sec:proppower}

Inserting eq.~(\ref{approxGret}) into eq.~(\ref{G_a}) we have
\begin{multline} \label{constantpower}  
  \mathcal{P}_p = \mathcal{P}_p^0
   \int_{-\infty}^{0}\!\!d\eta_1 \int_{-\infty}^{0}d\eta_2 \,  
   \frac{1}{p} \cos(p \eta_1) \cos(p \eta_2)
\\
  \times 
  e^{-\frac{1}{n + 1} \left(\frac{\gamma_p(\eta_1)}{H a(\eta_1)}
  + \frac{\gamma_p(\eta_2)}{H a(\eta_2)}\right)}
  N_p(\eta_1,\eta_2) 
\end{multline}
The factor $a^{-2}(\eta_{\rm fin})$ coming from the
rescaling of $\phi$  is compensated by the two factors 
of $a(\eta_{\rm fin})$ generated by the growing modes of $\chi_p$.
The integrand is now independent of $\eta_{\rm fin}$.

In the vacuum, i.e. at vanishing temperature, it has been shown by
general arguments~\cite{Parentani:2007uq,Parentani:2007dw} 
that the power spectrum agrees with the standard
prediction~(\ref{standardpower}) for $\Lambda \gg H$.
In other words, the double integral on the r.h.s.~of eq.~(\ref{constantpower})
evaluates to unity in this double limit. This can be shown by a lengthy
calculation which will be omitted here. We only state that
a naive analytic estimate can be found~\cite{Adamek2008} for the magnitude
of the leading order modification:
\begin{equation} \label{robustness}
   \frac{d \ln \delta\mathcal{P}_p}{d\ln(H/\Lambda)} = 
   n + \mathcal{O}\left( \frac{H^n}{\Lambda^n}  \right)\,,
\end{equation}
with $\delta\mathcal{P}_p~\equiv~\mathcal{P}_p - \mathcal{P}_p^0$.

\subsection{High temperature limit}
\label{sec:highT}

In keeping with the general character of our model, we do not specify
a physical motivation for the case of finite temperature, where 
$\Psi$ acts as a ``heat bath''. However, it serves to illustrate
the mechanism how 
the state of the adiabatic modes is dynamically 
determined by the state of the $\Psi$. In particular, if the latter are
in a thermal state, $\Phi$  
thermalizes through the interaction.
Furthermore, we establish that the thermal excitations of $\Phi_\pvec$ are
effectively populated at $\eta = \etaI$ 
defined by eq.~(\ref{Icrossing}). 

Let us consider the high temperature limit of eq.~(\ref{constantpower}),
i.e.~we insert
the limiting expression~(\ref{NhighT}) for
the noise kernel 
and set the $\delta$-function against one of the integrations. We then have
\begin{multline} \label{highTpspec}
   \mathcal{P}_p(T \to \infty)~=~
\mathcal{P}_p^0 \\\times 4 T
   \int_{-\infty}^{0}\!\!d\eta_1 \,
\frac{\gamma_p(\eta_1)}{p} a(\eta_1) \cos^2(p \eta_1)
  e^{-\frac{2}{n + 1} \frac{\gamma_p(\eta_1)}{H a(\eta_1)}}\,.
\end{multline}
Assuming that $\Lambda \gg H$, we may replace the squared cosine by
its average value $1/2$. Since $\gamma_p \propto a^{-n}$ in the relevant domain,
the remaining integral is now essentially a representation of the
Gamma function. It evaluates to
\begin{equation}
\label{highTresult}
   \mathcal{P}_p(T \to \infty) ~=~ 
   \mathcal{P}_p^0 \times \frac{T}{\Lambda} 
    \left(\frac{\Lambda}{H} \frac{2}{n + 1}\right)^{\frac{1}{n + 1}} 
    \Gamma\left(\frac{n}{n + 1}\right)~.
\end{equation}

If we compare this result to 
the high temperature limit of a thermal power spectrum
\begin{equation} \label{thermal}
\mathcal{P}_p(T)~=~\mathcal{P}_p^0 \times \coth\left(\frac{\Omega^\star}{2 T}\right)~,
\end{equation}
where $\Omega^\star$ denotes the proper frequency at the instant when the
occupation numbers are fixed, we find that
\begin{equation} \label{omegastar}
   \frac{\Omega^\star}{\Lambda}~=~\frac{2}{\Gamma\left(\frac{n}{n + 1}\right)} 
   \left(\frac{n + 1}{2} \frac{H}{\Lambda}\right)^{\frac{1}{n + 1}}~.
\end{equation}
One verifies that $\Omega^\star$ so defined coincides with the proper frequency at $\eta^\star_p$:
\begin{equation}
\Omega^\star~\simeq~\frac{\omega_p(\etaI)}{a(\etaI)}~.
\end{equation}
Note that it is below the scale $\Lambda$ (see eq.~(\ref{omegastar})),
in agreement with the results of Sec.~\ref{sec:approxG_ret}.

In conclusion, the state of $\phi$ is  
inherited from that of 
$\Psi$ at the time $\etaI$.

\section{Numerical analysis} 
\label{sec:num}

We present a numerical scheme that  solves the double time integral of
eq.~(\ref{G_a}) by means of Monte Carlo integration.

\subsection{The numerical scheme}

The procedure includes the following steps:
\begin{enumerate}
\item We impose the inflationary background. Specifically,
de Sitter space and power law inflation are considered.
\item We specify the ``relative'' 
decay rate $\gamma_p / p$ in terms of a function of $\pphys / \Lambda$, 
see eq.~(\ref{gammaseries}).
\item The equation of motion eq.~(\ref{chieqom}) for $\chi_p\left(\eta\right)$ 
is solved numerically.
\item The integral $\mathcal{I}_p\left(\eta, \eta'\right)$ 
of eq.~(\ref{visibility}) governing the 
amount of dissipation is computed numerically.
\item The retarded Green function is constructed from these numerical solutions,
cf.~eq.~(\ref{gretexplicit}).
\item The power spectrum is calculated from eq.~(\ref{finalpower}).
The derivatives and the double integral are  
evaluated numerically. 
\end{enumerate}

To proceed
we first need to address two technical issues. 
The first concerns the singular behavior of the 
noise kernel, the second the asymptotic behavior of $\gamma_p$
in the ultraviolet.

\subsection{Handling the singular behavior of $N$}

The noise kernel (cf.~eq.~(\ref{Nexplicit})) is singular for equal times
and should be interpreted as a Cauchy 
principal value. Unfortunately, numerical integrators
are generically incapable of calculating 
principal values; 
they usually fail to achieve convergence within a finite
number of integrand evaluations.
However in our case, the integrand can be
rewritten in a regular form by a convenient change of variables. 

We first note that the integrand of the double integral in eq.~(\ref{G_a}) 
is symmetric
under the exchange $\eta_1 \leftrightarrow \eta_2$. 
Thus, if we make a change of variables to $\zeta \equiv \eta_1 + \eta_2$,
$\xi \equiv \eta_1 - \eta_2$, the integrand will 
be symmetric around $\xi = 0$, where it has a double pole. 
Following the general techniques in the calculus of generalized functions \cite{Jones1966}, 
the integrand may be regularized by multiplication with
$-\xi^{2}\partial_\xi \xi^{-1} = 1$
and performing an integration by parts in $\xi$.
The derivative now acts on a regular (the $\xi^2$ cancels the double pole)
and symmetric function. 
It thus gives an odd function in $\xi$ such that
the remaining singular factor in $\xi^{-1}$ may be lifted.
This way we obtain a regular integrand that is well-behaved 
within the entire domain of integration. In brief, we have
 \begin{widetext}
 \begin{eqnarray} \label{finalpower}
 \mathcal{P}_p\left(\eta_{\rm fin}\right)
 &=& \frac{1}{2 a^2\left(\eta_{\rm fin}\right)} \iint \frac{\dd \zeta \dd \xi}{2} \, \GretPhi\left(\eta_{\rm fin}, \frac{\zeta + \xi}{2}\right) \GretPhi\left(\eta_{\rm fin}, \frac{\zeta - \xi}{2}\right) N_p\left(\frac{\zeta + \xi}{2}, \frac{\zeta - \xi}{2}\right)
 \left( - \xi^2 \partial_\xi \xi^{-1} \right)\nonumber\\
 &=&\frac{1}{2 a^2\left(\eta_{\rm fin}\right)} \iint \frac{\dd \zeta \dd \xi}{2}\, 
  \partial_\xi\left[\GretPhi\left(\eta_{\rm fin}, \frac{\zeta + \xi}{2}\right) \GretPhi\left(\eta_{\rm fin}, \frac{\zeta - \xi}{2}\right) N_p\left(\frac{\zeta + \xi}{2}, \frac{\zeta - \xi}{2}\right) \xi^2\right] \xi^{-1}\, .
 \end{eqnarray}
 \end{widetext}

\subsection{The UV behavior of the coupling function}
\label{sec:coupling}

For numerical integration, the range of $\eta$ has
to be truncated somewhere in the remote past.
In order to guarantee a safe truncation, we impose that the integrand drops off
exponentially. Then the cutoff can be chosen in such a way that the
truncation error is negligible  w.r.t.~the numerical value of the integral.

The exponential behavior of the integrand is achieved by a suitable choice
of the decay rate $\gamma_p$.
One might be tempted to say that any positive $\gamma_p$ gives rise to
an exponential behavior, simply given by $\mathcal{I}_p$. However, the
effective frequency of the mode functions depends on the damping rate
as well, cf.~eq.~(\ref{chieqom}). If 
$\gamma_p$ is not bounded from above, we may  
run into some pathologies 
due to an unbounded overdamping 
in the UV. The reason is that dissipation is less effective 
in an overdamped situation. 
To see this, let us consider a classical oscillator
with constant frequency $\omega$ and damping rate $\gamma$ 
in the overdamped regime, i.e.~$\gamma > \omega$.
It has two decaying modes with the decay rates 
$\gamma \pm \sqrt{\gamma^2 - \omega^2}$. 
When $\gamma^2 \gg \omega^2$,
the slowly decaying mode is thus
$\propto \exp[{-\frac{\omega^2}{2 \gamma} \left(\eta - \eta_0\right)}]$. 

Returning to our model, assuming 
$\gamma_p/p = (p/a\Lambda)^n$ without higher order terms,
the growing WKB solution of eq.~(\ref{chieqom})
for $\gamma_p \gg \omega_p$ would lead to an overall asymptotic behavior
$\propto \exp({-\int_{\eta_{\rm in}}^\eta 
\frac{\omega_p^2}{2 \gamma_p\left(\eta'\right)} \dd \eta'})$.  
As the integrand drops off like $a^n$ in the remote past,
the integral is generally finite in the limit $\eta_{\rm in}
\rightarrow -\infty$.
In this case nothing guarantees that one can
disregard the damped initial correlator in eq.~(\ref{fullG_a})
and that the power spectrum does not
depend on the initial state of $\Phi$ at $\eta_{\rm in}$.

However, when $\Lambda \gg H$, the residual contribution 
of the decaying term in eq.~(\ref{fullG_a}) is strongly suppressed
($\mathcal{O}\left[\exp\left(-\Lambda/H\right)\right]$)
because of the dissipation between $\Lambda$-crossing and horizon exit,
where the mode is in the underdamped regime.
But if we want to neglect the contribution of this decaying term
for all values of the ratio $\Lambda / H$, we cannot work with
$\gamma_p / p = (P / \Lambda)^n$
(nor with a polynomial of finite order)
since it is not bounded from above.
We will instead work with
a decay rate that saturates in the UV.
We choose 
\begin{equation} \label{numericgamma}
\frac{\gamma_p}{p}~=~\kappa~\mathrm{tanh}\left(\left(\frac{p}{a \Lambda}\right)^n \kappa^{-1}\right)\, .
\end{equation}
as a simple realization of this property. The new parameter $\kappa$
was introduced such that it only appears in the subleading
terms and that the decay rate saturates in the UV when it reaches the value
$\gamma_p / p \simeq \kappa$.

\begin{figure}[tb]
\includegraphics[width=90mm]{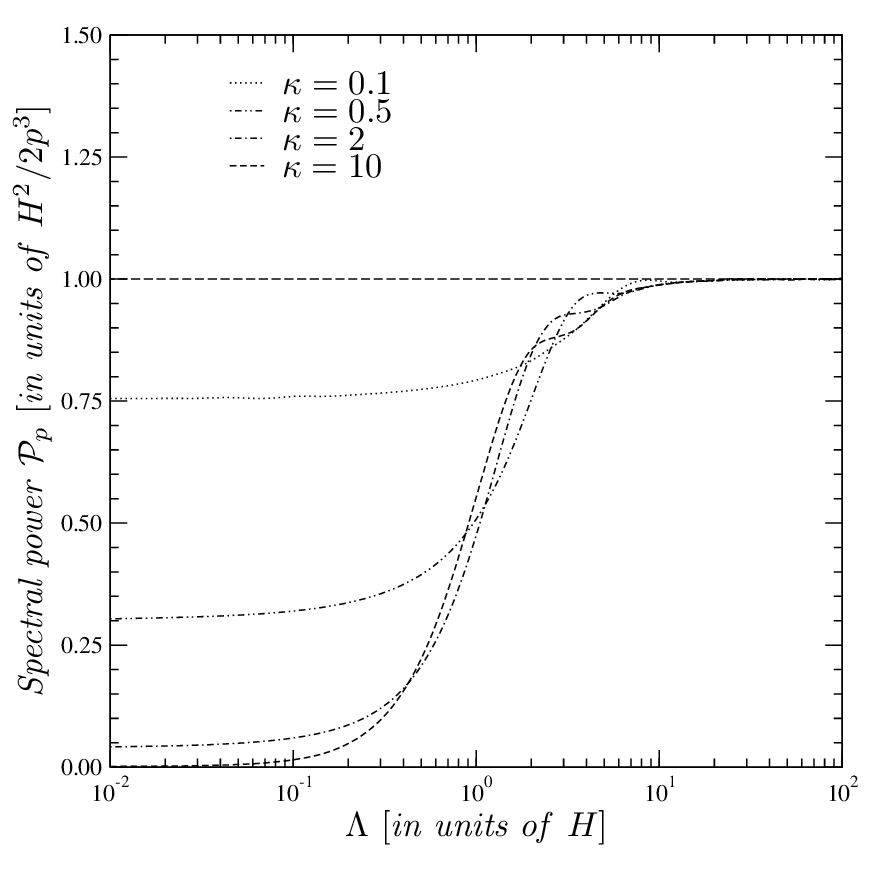}
\caption{\label{figdSall} \small 
Power spectrum 
$\mathcal{P}_p$ (in units of the standard power $\mathcal{P}_p^0$)
for $n = 2$ and $T = 0$ in de Sitter space, as a function of 
$\Lambda / H$ and for various choices of $\kappa$.
For $\Lambda \gg H$, the power spectrum asymptotes
from below to the standard value irrespectively
of the value of $\kappa$.
For $\Lambda \ll H$, the spectrum becomes flat again but 
it is lowered by a $\kappa$-dependent factor.
The region $\Lambda \lesssim H$ is non universal.}
\end{figure}

\begin{figure}[tb]
\includegraphics[width=90mm]{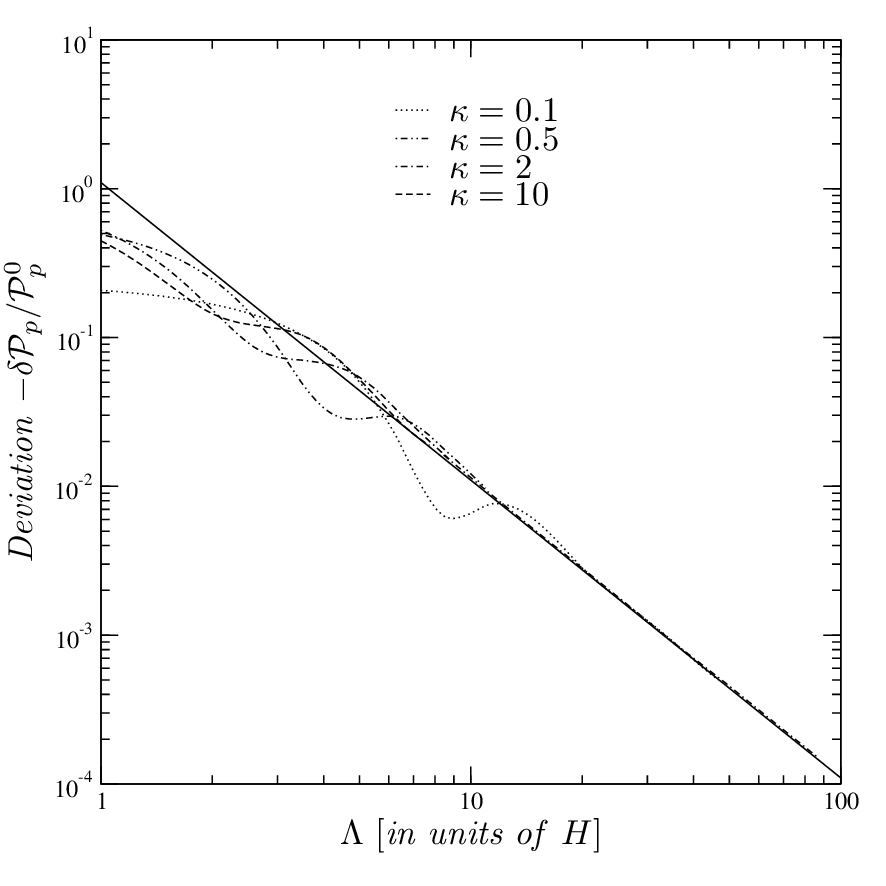}
\caption{\label{figdSpeak} \small Difference between the 
modified ($n = 2$, $T = 0$) and the standard power
spectrum in de Sitter space for $\Lambda \geq H$. 
The straight line indicates the first order deviation (which is
independent of $\kappa$) with a slope equal to $n$.} 
\end{figure}

\subsection{Numerical results}
\label{sec:results}

\subsubsection{Dependence on $H / \Lambda$ in the vacuum ($T = 0$)}
\label{sec:H/Lambda}

Let us now  consider the zero temperature limit, 
i.e. $\Psi$ is in its vacuum state.
On physical grounds, we expect that this will be the
relevant case if a fundamental theory gives rise to
dissipative effects in the UV sector
during inflation. 

\begin{figure}[tb]
\includegraphics[width=90mm]{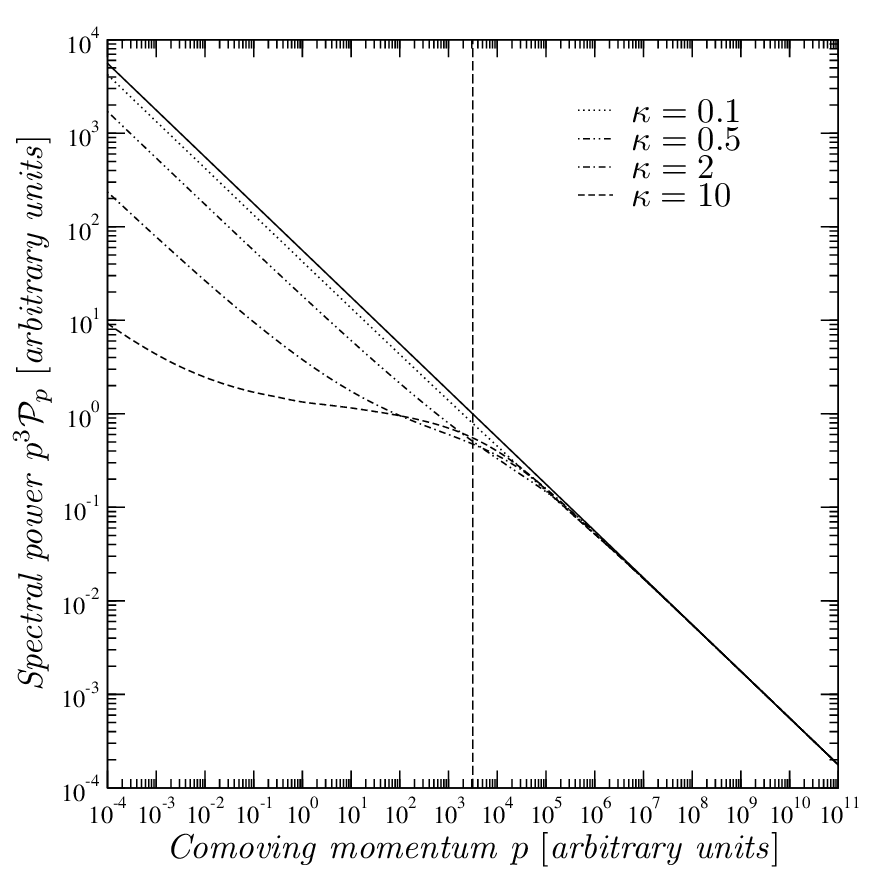}
\caption{\label{figpl} \small 
Power spectra for power law inflation ($\varepsilon = 0.2$). 
$n = 2$ and $T = 0$. The continuous 
line indicates the standard power $p^3 \mathcal{P}_p^0$.  
The vertical line
corresponds to the comoving momentum $p$ where $H_p = \Lambda$. 
For $H_p \ll \Lambda$ the power spectrum asymptotes to the standard value,
whereas in the other limit it is again sensitive to the UV.}
\end{figure}

The modified power spectrum is computed for $\Lambda$ ranging from
$\Lambda \gg H$ down  
to $\Lambda \ll H$ and is compared to the standard
prediction~(\ref{standardpower}),  
both in de Sitter space (figs.~\ref{figdSall} and~\ref{figdSpeak}),
and in power law inflation (fig.~\ref{figpl}). 

Let us first discuss the power spectrum in de Sitter space
as a function of $\Lambda$, fig.~\ref{figdSall}, coming from the high values.
Note that in de Sitter, $\Lambda / H$ is time independent, 
so the power spectrum is scale invariant and the value of
$\Lambda$ only affects the normalization of the spectrum.
As expected from the analytical results, for $\Lambda \gg H$ 
the standard power is recovered, independently of $\kappa$.
In this we corroborate the 
robustness of the power spectrum when the initial state is 
the adiabatic vacuum~\cite{NiemeyerParentani2001} 
(see also below in this section).
As $\Lambda$ approaches $H$, the power spectrum is modified
in a non-universal way which depends on all model parameters.
The flattening of the curves for $\Lambda \ll H$ has to be attributed to
the fact that our particular choice of $\gamma_p$ (cf.~eq.~(\ref{numericgamma}))
saturates to a constant in the UV and thus becomes independent of $\Lambda$
inside the horizon.

\begin{figure}[tb]
\includegraphics[width=90mm]{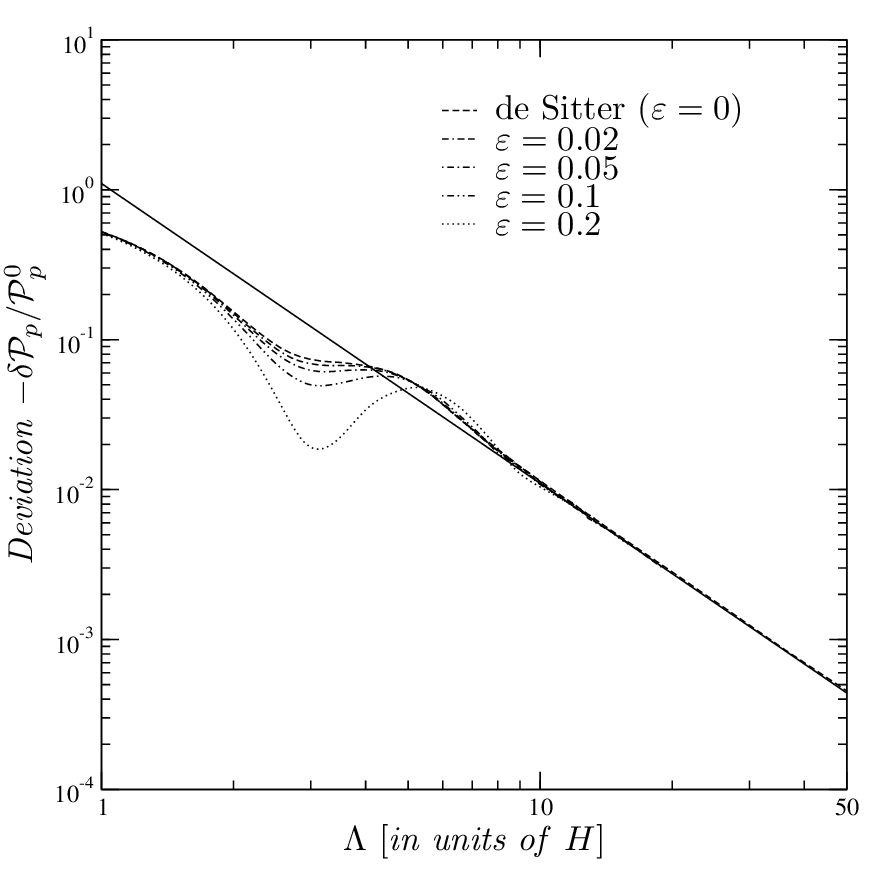}
\caption{\label{figplpeak} \small Difference between the modified 
($n = 2$, $\kappa = 2$, $T = 0$) and the standard power spectrum
for power law inflation and with $\Lambda \gtrsim H_p$.
Only the subleading modifications are sensitive to the value of
the slow-roll parameter $\varepsilon$.}
\end{figure}

Having established the robustness in the regime $H/\Lambda \ll 1$, 
the signature of dissipation is contained in the first deviation
with respect to the standard result. 
As anticipated in eq.~(\ref{robustness}),
in the limit $H/\Lambda \to 0$, 
the deviation behaves as
\begin{equation} \label{leading}
  \frac{\delta \mathcal{P}_p}{\mathcal{P}_p^0} \sim 
  \delta_n \times \left( \frac{H_p}
{\Lambda}\right)^n\, ,
\end{equation}
where the constant $\delta_n$ depends only on $n$.
This has been verified for values of $1 \leq  n \leq 2.5$,
as defined in eq.~(\ref{numericgamma}),
and we conjecture that it is valid for any power.
For all these values, $\delta_n$ is negative which means that the spectral power is reduced w.r.t.~the
standard value. We were not able to probe higher values of $n$
because of the numerical difficulties to follow the sharp decrease of the modifications.
However, preliminary results indicate that (as for dispersive
models~\cite{Macher:2008yq}) $\delta_n$ changes sign for $n = 3$, 
which means that for $n > 3$, dissipation leads to an increase of the power. 
At present we have no explanation for this unexpected result.

In fig.~\ref{figdSpeak}, the difference
$\delta \mathcal{P}_p = \mathcal{P}_p - \mathcal{P}_p^0$
is plotted for $n = 2$ and various values of $\kappa$
using a logarithmic scale.
It shows that the power spectrum becomes insensitive to the
next-to-leading order terms in the expansion of $\gamma_p$, 
see eq.~(\ref{gammaseries}). 

Let us now turn to power law inflation figs.~\ref{figpl} and~\ref{figplpeak}.
Fig.~\ref{figplpeak} indicates how the slow-roll parameter $\varepsilon$
impinges on the features of the modification $\delta\mathcal{P}_p$.
Surprisingly we find no dependence of $\delta_n$ on the slow-roll parameter, or if present,
it must be extremely mild. This implies that the power spectrum from power law inflation, as a 
function of $p$ (fig.~\ref{figpl}), is the power spectrum in de Sitter, as a function of
$H_p/\Lambda$,  where $H_p$ is the value of $H$ when the $p$ mode
exits the horizon, up to $\mathcal{O}(\gamma_p^2/p^2)$. 

Note that the modification of the power spectrum due to dissipative effects
can be reinterpreted in terms of a (scale dependent)
spectral index, which is given by the slope in fig.~\ref{figpl}.
To lowest order in $\varepsilon$ and $H_p/\Lambda$, 
using  eq. (\ref{leading}), we find
\begin{equation}
\frac{d \ln (p^3\mathcal{P}_p)}{d \ln p}~=~- 2\,\varepsilon
- \varepsilon~n~\delta_n \times \left(\frac{H_p}{\Lambda}\right)^n + \ldots~ 
\end{equation}
The first term accounts for the slow-roll evolution of the standard power spectrum,
whereas the second term is due to dissipative effects. It is proportional to $\varepsilon$
and suppressed by $(H_p/\Lambda)^n \ll 1$.

For $n = 2$, the negative sign of $\delta_n$ gives rise to a \textit{downward} modification of the
spectral power. Moreover, the effect is enhanced at large scales due to the
$p$-dependence of $H_p / \Lambda$.
The effect tends to increase the spectral index (i.e.~increase the slope at any given point of
the curve), and the running of the spectral index should indicate
a concave spectrum (the curve is bent downwards towards large scales).

\subsubsection{Dissipation and non-adiabaticity}

\begin{figure}[tb!]
\includegraphics[width=90mm]{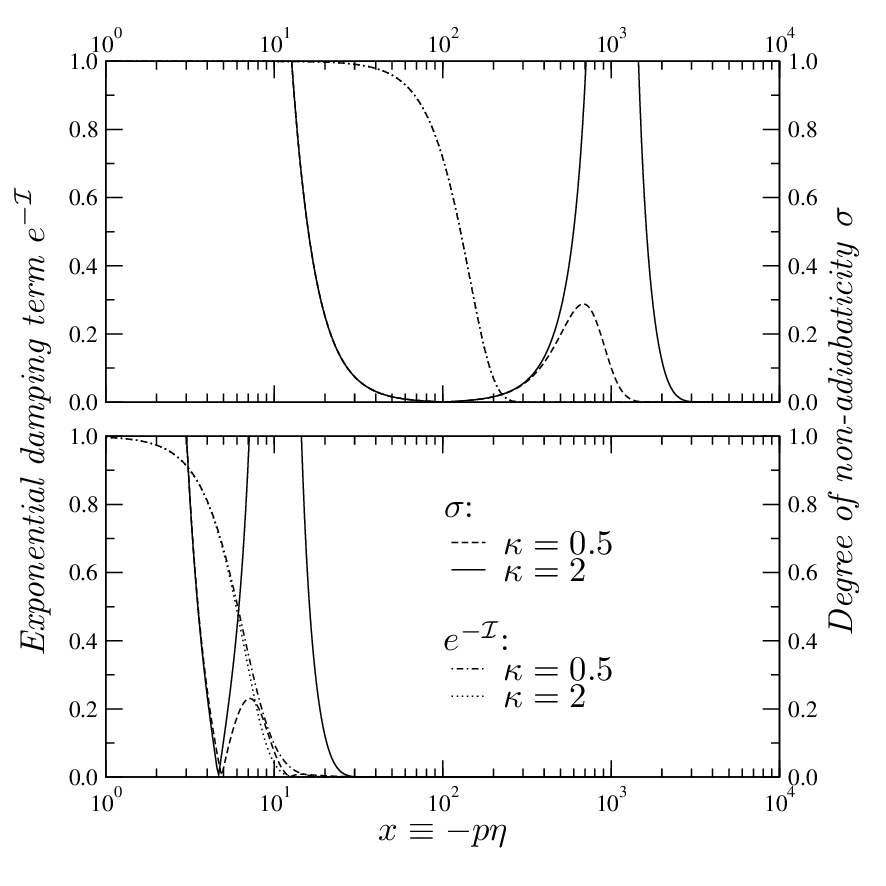}
\caption{\label{fig:nonadiabaticity-squar} \small 
The non-adiabaticity coefficient $\sigma$ and the exponential (\ref{visibility})
for $\kappa = 2$ and $\kappa = 0.5$. Upper panel: $\Lambda / H = 10^3$,
i.e.~the scales are clearly separated. Lower panel: $\Lambda / H = 10$, i.e.~weak scale
separation. The background is de Sitter, and $n = 2$. ($\sigma$ was enhanced by a factor
of $\Lambda / H$ for a clearer presentation.)}
\end{figure}

We now discuss the interplay between dissipation and 
non-adiabaticity in more detail.
Following what was done with dispersive models~\cite{Macher:2008yq},
we plot the function
\begin{equation}
   \sigma~=~\left| \frac{\partial_{\eta} 
   \omega_{\rm eff}}{\omega_{\rm eff}^2} \right|\, ,
\end{equation}
as a function of $x= -p\eta$ in fig.~\ref{fig:nonadiabaticity-squar},
where 
\begin{equation} \label{omegaeff}
   \omega_{\rm eff}^2 = \omega_p^2 - \gamma_p^2 \, ,
\end{equation} 
is the effective (conformal)
frequency of the modes $\chi_p$, see eq.~(\ref{chieqom}).
We see two regions of non-adiabaticity, the usual one at the horizon
$x \lesssim 1$, and a new UV feature,
inside the horizon, which is due to 
the dispersive effects induced by the dissipation and governed by $\gamma^2$. 
We also plot the exponential factor 
$\exp[-\mathcal{I}(0, \eta)]$, see eq.~(\ref{visibility}),
appearing in the retarded Green function.
We see that for sufficiently large $\Lambda$, the
UV feature in $\sigma(x)$ lies in the tail of the 
exponential. The non-adiabatic effects caused by dispersion
are then completely masked by dissipation.

As $\Lambda$ decreases, the UV feature in $\sigma(x)$ 
begins to overlap with the exponential.
For $\Lambda \simeq H$, it is well inside
the region where the exponential is still $\mathcal{O}(1)$.
In that case one expects
significant modifications of the power spectrum.

\subsubsection{Dependence on $T/\Lambda$}
\label{sec:T/Lambda}

\begin{figure}[tb]
\includegraphics[width=90mm]{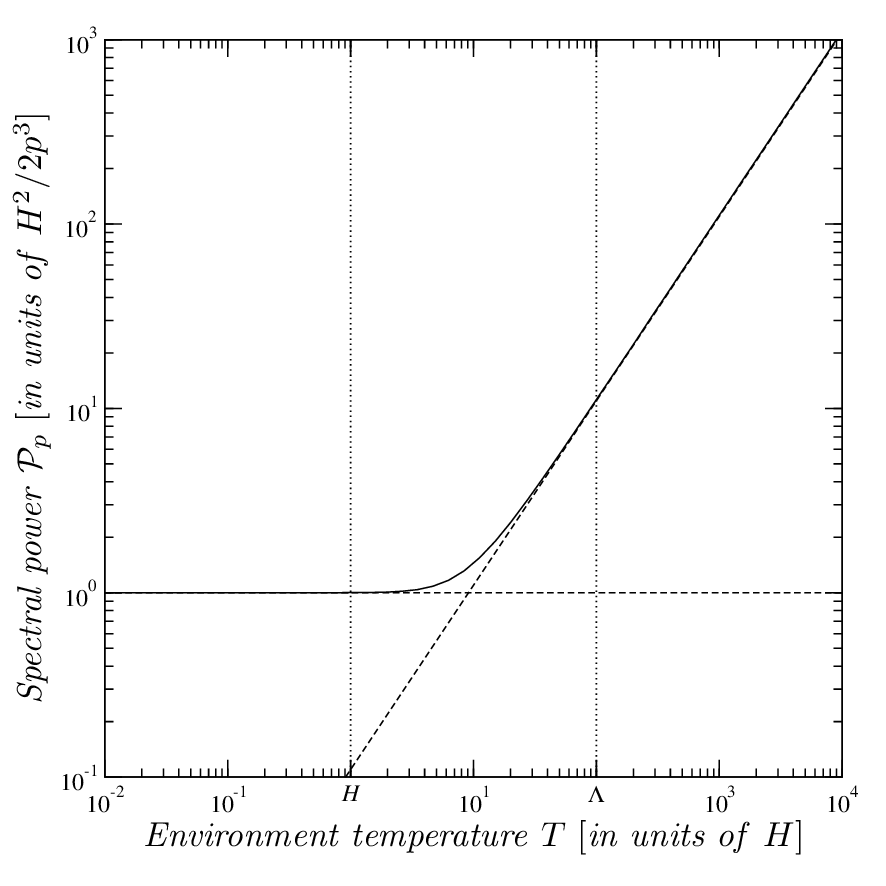}
\caption{\label{figdSvarT} \small Power spectrum 
as a function of the temperature $T$, computed for a
de Sitter background with $\Lambda = 100~H$ and $n = 2$. 
The result is therefore independent of $\kappa$.
For $T \ll H$, $\mathcal{P}_p$ asymptotes to $\mathcal{P}_p^0$ (horizontal dashed line), 
whereas for $T \gg H$ it asymptotes to the high temperature limit, eq.~(\ref{highTresult})
(the oblique dashed line).
The numerical result
neatly reproduces the power spectrum of a thermal state (cf.~eq.~(\ref{thermal}) and below)
whose temperature is specified at $\etaI$.}
\end{figure}

In this section we consider a de Sitter background and set $n=2$.
In fig.~\ref{figdSvarT} we plot the power spectrum as a function of $T$
in a case $H \ll \Lambda$.
For $T < H$ 
it asymptotes to the standard power $\mathcal{P}_p^0$, 
while for $T > \Lambda$ it 
converges to the analytic result~(\ref{highTresult}).
We verify that 
this curve is fitted at better than $1\%$ by the power spectrum 
obtained in non-dissipative and relativistic settings 
when the ``initial'' state of $\Phi$ 
is enforced 
to be that thermal state
at $\eta = \etaI$, 
see eq.~(\ref{thermal}).

\section{Conclusion}
\label{sec:discussion}

Let us sum up our main results.
First, in spite of the strong dissipative effects encountered in the
early mode propagation, the predictions for the power spectrum
converge to the standard ones
in the case of scale separation, $H_p \ll \Lambda$,
and if the environment field $\Psi$ is in the ground state.
The power spectrum is hence a robust observable with respect to
high-energy dissipative effects under these conditions.
In this regard, dissipative models do not differ from 
dispersive ones~\cite{NiemeyerParentani2001}.

Second, the leading deviation of the power spectrum induced 
by dissipation is linear in the relative decay rate $\gamma_p/p$ 
evaluated at horizon crossing, see eq.~(\ref{leading}). 
The signatures of dissipation therefore do not oscillate.
In this regard as well, dissipative models behave like 
dispersive ones~\cite{Macher:2008yq}. In the
region of parameter space we succeeded to explore,
i.e. the power $n \leq 2.5$,
the deviations are negative,
giving rise to a suppression of the power spectrum with respect to the
standard one.

Third, we have verified 
that the deviations in power law inflation essentially
behave as those evaluated in de Sitter space, with 
$H$ replaced by $H_p$ evaluated at horizon exit. 
As a direct consequence, any observable effects of high-energy
dissipation will be more pronounced at the largest accessible scales,
corresponding to the largest $H_p/\Lambda$.

Fourth, in the case of $H_p \ll \Lambda$, we  
showed how dissipation, via the function $\mathcal{I}_p$, 
sets the time when the ``initial'' state of the 
$\Phi_\pvec$ mode is effectively set. 
It is given by eq.~(\ref{Icrossing}) where $P^\star = p/a(\etaI)$
is at an intermediate scale between $H_p$ and $\Lambda$.
In addition, the state of $\Phi_\pvec$ coincides 
with that of the degrees of freedom causing the dissipation. 
Thus, the properties of the state of $\Psi_k$ pass on to that of $\Phi_\pvec$
at $\etaI$. 

Fifth, in Sec.~\ref{general}, we gave (sufficient)
conditions under which a model exhibiting
dissipation above the scale $\Lambda$
can be well approximated by a Gaussian model, in the sense that 
both models predict the same power spectrum.

Our analysis could be extended in two directions.
First, we considered a Gaussian model. If we relax this hypothesis, 
since dissipation grows with the coupling,
it will be interesting to investigate the
combined effects of dissipation and non-Gaussianities.
Second, we calculated $\mathcal{P}^{\phi}_p$, 
the power spectrum of a test field
propagating on an inflationary background. 
It is a challenge to construct a realistic model of inflation
displaying dissipation in the UV sector.  

We can nevertheless make the following observations.
At the linearized level,
the spectrum of scalar metric perturbations $\zeta$ is related to that of our scalar field $\phi$
by $\mathcal{P}^{\zeta}_p = \frac{4 \pi G}{\varepsilon_p} \mathcal{P}^{\phi}_p$~\cite{MukhPhysRep}.
This implies that the \textit{relative} modification $\delta  \mathcal{P}^{\zeta}_p / \mathcal{P}^{\zeta}_p$
due to some dissipative effects is the same as the one of our scalar field.
Therefore, if $\Psi$ couples identically to 
scalar and tensor perturbations, the $S/T$ ratio should not be changed at first order.
Then the absence of features in the 
power spectra in the regime $H/\Lambda \ll 1$ 
would prevent us from disentangling the new physics from a
simple shift of the inflaton potential, adding yet another ambiguity 
to the program of the inflaton potential reconstruction.

\acknowledgements

The work of DC and JCN was supported by the Alfried Krupp Prize for Young 
University Teachers of the Alfried Krupp von Bohlen und Halbach
Foundation. The numerical power spectra were computed using Monte Carlo
integrators from the \textsc{Cuba} package \cite{Hahn2005}. Special
thanks go to Alex Schenkel for interesting discussions and many helpful
comments. We also acknowledge contributions from Tim Koslowski, Dennis
Simon and many other colleagues.


\end{document}